\documentclass[jcp,onecolumn,superscriptaddress,12pt]{revtex4}
\usepackage{graphicx}
\usepackage{subfigure}
\usepackage{amsmath}
\usepackage{amsfonts}
\usepackage{amssymb}
\usepackage{amsthm}
\usepackage{times}
\usepackage[colorlinks,citecolor=blue,linkcolor=blue]{hyperref}
\usepackage{color}
\usepackage{setspace}

\begin{document}
\title{Efficient and Deterministic Propagation of Mixed Quantum-Classical Liouville Dynamics}

\newcommand{\UA}{\affiliation{Department of Chemistry, University of Alberta, Edmonton, Alberta T6G 2G2, Canada}}

\date{\today}
\author{Junjie Liu}
\UA
\author{Gabriel Hanna}
\email{gabriel.hanna@ualberta.ca}
\UA

\begin{abstract}
We propose a highly efficient mixed quantum-classical molecular dynamics scheme based on a solution of the quantum-classical Liouville equation (QCLE). By casting the equations of motion for the quantum subsystem and classical bath degrees of freedom onto an approximate set of coupled first-order differential equations for {\it c}-numbers, this scheme propagates the composite system in time deterministically in terms of independent classical-like trajectories. To demonstrate its performance, we apply the method to the spin-boson model, a photo-induced electron transfer model, and a Fenna-Matthews-Olsen complex model, and find excellent agreement out to long times with the numerically exact results, using several orders of magnitude fewer trajectories than surface-hopping solutions of the QCLE.  Owing to its accuracy and efficiency, this method promises to be very useful for studying the dynamics of mixed quantum-classical systems.
\end{abstract}

\maketitle

The computational study of quantum dynamical processes occurring in condensed phase environments often requires an accurate treatment of the coupling between the subsystem of primary interest and its environment.  For instance, the rates and mechanisms of transfer processes involving protons, electrons, excitonic energy, and quantum states from a donor to an acceptor are often influenced by the fluctuations in their environments.  However, a fully quantum dynamical simulation of a system undergoing such a process is prohibitively expensive due to the large number of degrees of freedom (DOFs) in the environment.  In many such cases, mixed quantum-classical methods,\cite{Tully.90.JCP,Billing.93.JCP,Prezhdo.97.PRA,Martens.97.JCP,Donoso.98.JPCA,Tully.98.FD,Kapral.99.JCP,Donoso.00.JCP,Wan.00.JCP,Horenko.02.JCP,Wan.02.JCP,Horenko.04.JCP,Roman.07.JPCA,Kelly.13.JCP,Bai.14.JPCA,Kim.14.JCP,Kim.14.JCPa,Wang.15.JPCL,Martens.16.JPCL,Wang.16.JPCL,Agostini.16.JCTC,Subotnik.16.ARPC} which treat the subsystem of interest quantum mechanically and its environment in a classical-like fashion, constitute attractive alternatives to fully quantum mechanical ones.  

The quantum-classical Liouville equation (QCLE) \cite{Aleksandrov.81.ZNA,Gerasimenko.82.TMP,Zhang.88.JPP,Kapral.99.JCP} has given rise to arguably the most rigorous mixed quantum-classical dynamics algorithms to date.  If solved exactly, the QCLE can even reproduce the exact quantum dynamics of arbitrary quantum subsystems that are bilinearly coupled to harmonic environments \cite{Kernan.02.JCP},  which are commonly encountered in chemical physics.  However, the approximations and/or instabilities inherent to the previous algorithms for solving the QCLE \cite{Kernan.02.JPCM,Hanna.05.JCP,Kernan.08.JPCB,Kim.08.JCP,Hsieh.12.JCP,Kapral.15.JP,Kananenka.16.JPCL} have restricted their broad-scale applicability. In particular, the surface-hopping solutions \cite{Kernan.02.JPCM,Hanna.05.JCP,Kernan.08.JPCB} suffer from numerical instabilities induced by a Monte Carlo sampling of the nonadiabatic transitions and, consequently, require very large ensembles of trajectories for convergence of the results.  On the other hand, the mapping-basis solutions \cite{Kim.08.JCP,Hsieh.12.JCP} require much smaller ensembles of trajectories, but they can yield unsatisfactory results in certain situations due to their inherent mean-field-like approximations.

Our goal is to demonstrate that, after making a series of assumptions, one can simulate the coupled subsystem-environment dynamics resulting from the QCLE with high accuracy, high stability, and low computational cost. The scheme proposed herein allows one to compute the expectation values of time-dependent observables using deterministic, independent, and classical-like molecular dynamics (MD) trajectories.  In contrast to the surface-hopping solutions of the QCLE \cite{Kernan.02.JPCM,Hanna.05.JCP,Kernan.08.JPCB}, this scheme represents both the quantum and classical DOFs in terms of continuous variables and does not involve stochastic hops between potential energy surfaces.  As will be shown, our approach provides an effective way of simulating the dynamics of mixed quantum-classical systems.  

We start by introducing the Weyl-ordered, partially-Wigner transformed Hamiltonian that governs the QCL dynamics of a system 
\begin{equation}\label{eq:htot_w}
\hat{H}_W~=~\hat{H}_S(\boldsymbol{\hat{x}})+H_B(\boldsymbol{X})+\hat{V}_C(\boldsymbol{\hat{x}},\boldsymbol{X}),
\end{equation}
where $\hat{H}_S$ is the quantum subsystem Hamiltonian of dimensionality $L$ and $\boldsymbol{\hat{x}}=(\hat{x}_1,\hat{x}_2,\cdots,\hat{x}_{L^2-1})$ denotes a set of generalized coordinates that provides a complete description of the state of the subsystem. For example, one could choose $\boldsymbol{\hat{x}}=(\hat{\sigma}_x,\hat{\sigma}_y,\hat{\sigma}_z)$ for a two-level spin subsystem (where $\hat{\sigma}_{x/y/z}$ denote the Pauli matrices) or, more generally, projection operators for multi-level subsystems \cite{Hioe.81.PRL}. The set contains $L^2-1$ coordinates because there are $L^2-1$ independent elements in the reduced density matrix $\hat{\rho}_S$, which is Hermitian and satisfies $\mathrm{Tr}_S\hat{\rho}_S=1$. $H_B$ is the bath (or environment) Hamiltonian, where $\boldsymbol{X}=(\boldsymbol{R},\boldsymbol{P})$ with $\boldsymbol{R}=(R_1,R_2,\cdots,R_N)$ and $\boldsymbol{P}=(P_1,P_2,\cdots,P_N)$, and $\hat{V}_C$ denotes the subsystem-bath coupling potential.  Weyl ordering (e.g., a product term $\boldsymbol{\hat{x}}\boldsymbol{X}$ would be rewritten as $(\boldsymbol{\hat{x}}\boldsymbol{X}+\boldsymbol{X}\boldsymbol{\hat{x}})/2$) is required to account for the noncommutativity of the subsystem and bath coordinates in our new scheme, as will be elaborated upon below. The subscript $W$ indicates that the partial Wigner transform over the bath DOF has been taken. 

The basis-free QCLE for an arbitrary observable $\hat{A}$ of this system, expressed in the Eulerian frame  of reference (i.e., the dynamics is viewed at a fixed point $\boldsymbol{X}$), is given by \cite{Kapral.99.JCP}
\begin{eqnarray}\label{eq:QCLE}
\frac{\partial}{\partial t}\hat{A}_W(\boldsymbol{X},t)~&=&~\frac{i}{\hbar}[\hat{H}_W(\boldsymbol{X}),\hat{A}_W(\boldsymbol{X},t)]-\{\hat{H}_W(\boldsymbol{X}),\hat{A}_W(\boldsymbol{X},t)\}_a\nonumber \\ 
&\equiv& i\hat{\mathcal{L}}_W\hat{A}_W(\boldsymbol{X},t),
\end{eqnarray}
where $\{\cdot,\cdot\}_a$ is the anti-symmetrized Poisson bracket, namely $\{\hat{H}_W,\hat{A}_W\}_a=\frac{1}{2}\{\hat{H}_W,\hat{A}_W\}-\frac{1}{2}\{\hat{A}_W,\hat{H}_W\}$, and the second line of this equation defines the QCL operator $\hat{\mathcal{L}}_W$. A number of numerical methods for solving the QCLE, which differ in the basis chosen to represent the quantum subsystem operators, have been developed \cite{Wan.00.JCP,Horenko.02.JCP,Wan.02.JCP,Horenko.04.JCP,Kernan.02.JPCM,Hanna.05.JCP,Kernan.08.JPCB,Kim.08.JCP,Hsieh.12.JCP}.  However, these methods have been shown to be either limited by their underlying approximations or their high computational costs.
 
Instead of propagating the observable directly as in the previous methods, our new algorithm computes the time dependence of $\hat{A}_W(t)$ from the dynamics of the coordinates $\boldsymbol{\hat{x}}(t)$ and $\boldsymbol{X}(t)$, starting from a factorized initial state $\hat{\rho}_W(0)=\hat{\rho}_S(0)\rho_{B,W}(0)$. To obtain $\boldsymbol{\hat{x}}(t)$ and $\boldsymbol{X}(t)$, one must move to the Lagrangian frame of reference, in which the quantum subsystem evolves in time along with the classical phase space coordinates. As the partial Wigner transform introduced above was performed with respect to the initial phase space point $\boldsymbol{X}$, one cannot directly apply Eq.~(\ref{eq:QCLE}) to obtain $\boldsymbol{\hat{x}}(t)$ and $\boldsymbol{X}(t)$ in the Lagrangian frame.  Rather, according to Eq.~(\ref{eq:QCLE}), the subsystem and bath coordinates satisfy $\left.\boldsymbol{\dot{\hat{x}}}\right|_{t=0}~=~\frac{i}{\hbar}[\hat{H}_W, \boldsymbol{\hat{x}}]$, $\left.\boldsymbol{\dot{X}}\right|_{t=0}~=~-\{\hat{H}_W,\boldsymbol{X}\}_a$ (where the dot denotes a time derivative).  However, one can show that if only zeroth-order terms in $\hbar$ are retained in the Moyal product between $e^{i\hat{\mathcal{L}}_Wt}$ and an arbitrary operator $\hat{B}$ at finite times, i.e., 
\begin{equation}
(\hat{B}(\boldsymbol{\hat{x}}(t),\boldsymbol{\hat{X}}(t)))_W~\approx~(\hat{B}_W(\boldsymbol{\hat{x}},\boldsymbol{X}))(t), 
\end{equation}
one may generalize the equations of motion (EOMs) of the coordinates at the initial time to finite times (see section I of the Supporting Information (SI) for the details of how this is done and the assumptions involved), namely
\begin{equation}\label{eq:eom_general}
\boldsymbol{\dot{\hat{x}}}(t)~=~\frac{i}{\hbar}\left([\hat{H}_W, \boldsymbol{\hat{x}}]\right)(t),~~~\boldsymbol{\dot{X}}(t)~=~-\left(\{\hat{H}_W, \boldsymbol{X}\}_a\right)(t).
\end{equation}
In the above equation, the time arguments are placed outside of their respective brackets to indicate that one should first evaluate the commutator and Poisson brackets with respect to the initial bath coordinates (in accordance with the partial Wigner transform) and then apply the time dependence to the coordinates in the resulting expressions.

The next step is to cast Eq.~(\ref{eq:eom_general}) in an arbitrary basis $\{|\alpha\rangle\}=(|\alpha_1\rangle,\ldots,|\alpha_L\rangle)$ that spans the Hilbert space of the $L$-dimensional quantum subsystem (the exact nature of this basis would be chosen based on convenience). For example, the EOMs for the matrix elements of $\boldsymbol{\hat{x}}(t)$ and $\boldsymbol{X}(t)$ for a subsystem that is bilinearly coupled to a harmonic bath are 
\begin{eqnarray}\label{eq:eom_detail}
\boldsymbol{\dot{x}}^{\alpha\alpha^{\prime}}(t) &=& F(\{\boldsymbol{x}^{\alpha\alpha^{\prime}}(t)\},\{(\boldsymbol{\hat{x}}(t)\boldsymbol{X}(t)+\boldsymbol{X}(t)\boldsymbol{\hat{x}}(t))^{\alpha\alpha^{\prime}}\}),\nonumber\\
\boldsymbol{\dot{X}}^{\alpha\alpha^{\prime}}(t) &=& G(\{\boldsymbol{x}^{\alpha\alpha^{\prime}}(t)\},\{\boldsymbol{X}^{\alpha\alpha^{\prime}}(t)\}),
\end{eqnarray}
where $D^{\alpha\alpha^{\prime}}\equiv\langle\alpha|D|\alpha^{\prime}\rangle$, $F\equiv\frac{i}{\hbar}\langle\alpha|\left([\hat{H}_W, \boldsymbol{\hat{x}}]\right)(t)|\alpha^{\prime}\rangle$ is a functional of the matrix elements $\boldsymbol{x}^{\alpha\alpha^{\prime}}(t)$ and $(\boldsymbol{\hat{x}}(t)\boldsymbol{X}(t)+\boldsymbol{X}(t)\boldsymbol{\hat{x}}(t))^{\alpha\alpha^{\prime}}$ (which arises from the bilinear interaction in the Weyl-ordered Hamiltonian $\hat{H}_W$), and $G\equiv-\langle\alpha|\left(\{\hat{H}_W, \boldsymbol{X}\}_a\right)(t)|\alpha^{\prime}\rangle$ is a functional of the matrix elements $\boldsymbol{x}^{\alpha\alpha^{\prime}}(t)$ and $\boldsymbol{X}^{\alpha\alpha^{\prime}}(t)$. In the above, the notation $\{\boldsymbol{z}^{\alpha\alpha^{\prime}}\}$ denotes a particular set of matrix elements of $\boldsymbol{z}$ in the basis $\{|\alpha\rangle\}$ (the contents of which depend on the model under investigation). The detailed forms of $F$ and $G$ must be worked out for the system under study (e.g., the explicit forms of $F$ and $G$ for the models considered in this work are shown in the SI). It should be noted that the superscript in $\boldsymbol{X}^{\alpha\alpha^{\prime}}(t)$ serves as a label to distinguish the various $\it{c}$-numbers (and their corresponding EOMs) that arise due to the subsystem-bath coupling. Since $(\hat{x}_lX_k)^{\alpha\alpha^{\prime}}=\sum_{\beta}x_l^{\alpha\beta}X_k^{\beta\alpha^{\prime}}$, one may interpret Eq.~(\ref{eq:eom_detail}) as a set of coupled first-order differential equations (FODEs) for the {\it c}-numbers ($\boldsymbol{x}^{\{\alpha\alpha^{\prime}\}}(t),\boldsymbol{X}^{\{\alpha\alpha^{\prime}\}}(t)$), where $\{\alpha\alpha^{\prime}\}$ denotes all the combinations of basis indices. The maximum number of coupled FODEs is $L^2(L^2-1+2N)$ (because one could reduce this number if the subsystem has symmetry). 
 
Within the QCL formalism, the expectation value of an observable $\hat{A}$ can be expressed as $\langle \hat{A}(t) \rangle = \sum_{\alpha\alpha^{\prime}}\int d\boldsymbol{X}A_W^{\alpha\alpha^{\prime}}(\boldsymbol{\hat{x}},\boldsymbol{X},t)\rho_W^{\alpha^{\prime}\alpha}(\boldsymbol{\hat{x}},\boldsymbol{X})$, where $\rho_W^{\alpha^{\prime}\alpha}(\boldsymbol{\hat{x}},\boldsymbol{X})$ denotes a matrix element of the partially Wigner transformed initial total density operator \cite{Sergi.03.TCA}.  Based on this expression, one can write down the following rule for constructing the time-dependent expectation value of an observable in terms of the time-dependent {\it c}-numbers: 
\begin{eqnarray}\label{eq:average}
\langle \hat{A}(t) \rangle &=& \sum_{\alpha\alpha^{\prime}}\int d\boldsymbol{X}(0)A_W^{\alpha\alpha'}(\boldsymbol{\hat{x}}(t),\boldsymbol{X}(t)) \rho_W^{\alpha^{\prime}\alpha}(\boldsymbol{\hat{x}}(0),\boldsymbol{X}(0)).
\end{eqnarray}
To execute the above rule, one must first specify the initial values of the matrix elements $\boldsymbol{x}^{\{\alpha\alpha^{\prime}\}}(0)$ and $\boldsymbol{X}^{\{\alpha\alpha^{\prime}\}}(0)=\boldsymbol{X}(0)\delta_{\{\alpha\alpha^{\prime}\}}$.  For the factorized initial state, i.e., $\rho_W^{\alpha^{\prime}\alpha}(\boldsymbol{\hat{x}}(0),\boldsymbol{X}(0))=\rho_{B,W}(\boldsymbol{X}(0)) \rho_S^{\alpha^{\prime}\alpha}(\boldsymbol{\hat{x}}(0))$, $\boldsymbol{x}^{\{\alpha\alpha^{\prime}\}}(0)$ is determined after specifying the basis and $\boldsymbol{X}(0)$ is sampled from $\rho_{B,W}(\boldsymbol{X}(0))$. Then, for each set of initial conditions, one uses a numerical integration scheme such as the Runge-Kutta method \cite{Dormand.80.JCAM} to integrate the $L^2(L^2-1+2N)$ coupled FODEs [i.e., Eq.~(\ref{eq:eom_detail})] up to time $t$.  Using the resulting $(\boldsymbol{x}^{\{\alpha\alpha^{\prime}\}}(t),\boldsymbol{X}^{\{\alpha\alpha^{\prime}\}}(t))$, one evaluates the required terms in the summand and integrand of Eq.~(\ref{eq:average}).  Finally, one averages over an ensemble of trajectories to compute $\langle \hat{A}(t)\rangle$.  Together, Eqs.~(\ref{eq:eom_detail}) and (\ref{eq:average}) prescribe a deterministic, classical-like MD scheme for simulating the time evolution of observables in mixed quantum-classical systems.  If one would like to calculate quantum equilibrium correlation functions, the above construction rule would change, but the spirit of the approach would remain the same.  In light of the nature of our QCLE-based method, we will refer to it as DECIDE (i.e., \textbf{D}eterministic \textbf{E}volution of \textbf{C}oordinates with \textbf{I}nitial \textbf{D}ecoupled \textbf{E}quations).

The DECIDE method has a number of advantages over existing mixed quantum-classical approaches: (i) The time evolution prescribed by the FODEs is deterministic, which results in numerically stable results out to long times. For the models considered in this work, ensembles of only a few thousand trajectories suffice to obtain well-converged results, compared to the, at least, $10^5-10^6$ trajectories required by the other QCLE-based methods. (ii) The scaling of this method is polynomial in $L$ and $N$, as it only requires the integration of at most $L^2(L^2-1+2N)$ coupled FODEs. (iii) There is no need to diagonalize the Hamiltonian matrix on-the-fly as in the surface-hopping methods (see section II of the SI for an elaboration on this point). (iv) This method does not rely on the momentum jump approximation \cite{Sergi.03.TCA,Hanna.05.JCP}, which is required to obtain a surface-hopping solution of the QCLE. (v) The time evolution prescribed by the FODEs is not of a mean-field type. In contrast to Ehrenfest dynamics, where the classical coordinates feel an average force determined by the total wave function of the quantum subsystem, this method involves a set of equations of motion for a given $(R_j,P_j)$ whose individual equations differ from one another due to their dependencies on different subsystem matrix elements (and therefore involve different state-dependent forces). 

To illustrate the use of DECIDE, we apply it to three models:  the spin-boson model (SBM) \cite{Leggett.87.RMP,Weiss.12.NULL}, a photo-induced electron transfer (PIET) model \cite{Wang.04.CPL}, and a Fenna-Matthews-Olsen (FMO) complex model \cite{Fenna.75.N,Adolphs.06.BJ,Ishizaki.09.PNAS}.  We assess its performance by comparing our results to numerically exact benchmarks.    

We start by considering the unbiased SBM, whose Weyl-ordered Hamiltonian takes the form 
\begin{equation}
\hat{H}_{W}=-\hbar\Delta \hat{\sigma}_x+\frac{1}{2}\sum_{j=1}^N\left(P_j^2+\omega_j^2R_j^2-C_jR_j\hat{\sigma}_z-C_j\hat{\sigma}_zR_j\right),
\end{equation}
where $\hat{\sigma}_{x/z}$ are the Pauli spin matrices, $\Delta$ is the tunneling frequency between spin states, $\omega_j$ is the frequency of the $j$th harmonic oscillator, $C_j$ is the coupling coefficient between the spin and the $j$th harmonic oscillator, and $N$ is the number of harmonic oscillators. The bilinear subsystem-bath coupling is characterized by an Ohmic spectral density with an exponential cutoff, namely $J(\omega)=\frac{\xi}{2}\pi\omega e^{-\omega/\omega_c}$, where the Kondo parameter $\xi$ characterizes the subsystem-bath coupling strength and $\omega_c$ is the cut-off frequency.  

For this model, the three Pauli matrices are chosen as the generalized subsystem coordinates, i.e., $\boldsymbol{\hat{x}}=(\hat{\sigma}_x,\hat{\sigma}_y,\hat{\sigma}_z)$.  Therefore, Eq.~(\ref{eq:eom_detail}) consists of $4\times(3+2N)$ coupled FODEs for the matrix elements of the subsystem and bath coordinates. The initial state is given by $\hat{\rho}_W(0)=\rho_{B,W}(0)\hat{\rho}_S(0)$, where $\hat{\rho}_S(0)=|+\rangle\langle+|$ (with $|\pm\rangle$ defined by $\hat{\sigma}_z|\pm\rangle=\pm|\pm\rangle$) and $\rho_{B,W}(0)=\prod_{j=1}^N\frac{\tanh(\hbar\beta\omega_j/2)}{\pi}\exp\left[-\frac{2\tanh(\hbar\beta\omega_j/2)}{\hbar\omega_j}\left(\frac{P_j^2}{2}+\frac{\omega_j^2R_j^2}{2}\right)\right]$ (with the inverse temperature $\beta$). Given the form of $\hat{\rho}_S(0)$,  we choose $\{|\alpha\rangle\}=\{|+\rangle, |-\rangle\}$; thus, according to Eq.~(\ref{eq:average}), the expectation value of the spin population difference is 
\begin{equation}
\langle\hat{\sigma}_z (t)\rangle~=~\int d\boldsymbol{X}(0)\rho_{B,W}(\boldsymbol{X}(0))\sigma_z^{++}(t),
\end{equation}
where we have used the fact that $\rho_S^{++}(0)=1$.

\begin{figure}[tbh!]
  \centering
  \includegraphics[scale=0.85]{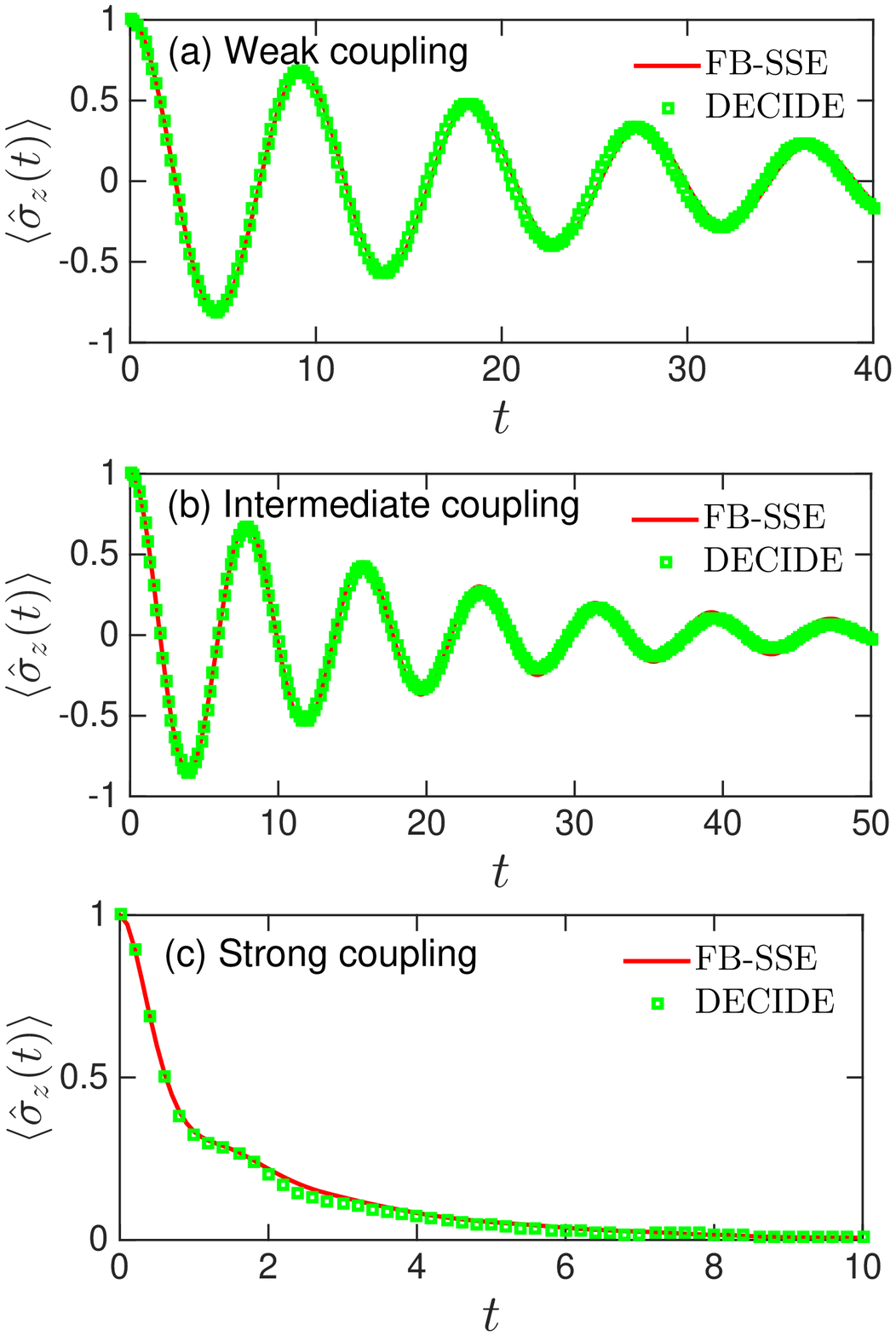}
\caption{Time evolution of $\langle\hat{\sigma}_z(t)\rangle$ for the spin-boson model in (a) the weak coupling regime with $\xi=0.007$, $\Delta=1/3$, and $\beta=0.3$, (b) the intermediate coupling regime with $\xi=0.09$, $\Delta=0.4$, and $\beta=12.5$, and (c) the strong coupling regime with $\xi=2$, $\Delta=1.2$, and $\beta=0.25$. An ensemble of $1\times10^4$ trajectories and a MD time step of $\Delta t=0.02$ were used to obtain our converged DECIDE results (green squares). The red solid lines are the benchmark results calculated using the FB-SSE method. The values of the remaining parameters are $\omega_c=1$, $\omega_{max}=5$, and $N=100$.}
\label{fig:pop_sb}
\end{figure}

Our results for $\langle\hat{\sigma}_z(t)\rangle$ in the weak, intermediate, and strong coupling regimes are shown in figure \ref{fig:pop_sb} (the simulation details may be found in section II of the SI).
The benchmark results were obtained using a numerically exact method known as the forward-backward stochastic Schr\"odinger equation (FB-SSE) \cite{Ke.16.JCP}. As can be seen, the DECIDE results, generated using only $1\times10^4$ trajectories, are in excellent agreement with the benchmark results out to long times. (It should be noted that reasonable results can already be obtained with as few as $1\times 10^3$ trajectories.) These results should be contrasted with those obtained by one of the authors using a surface-hopping solution of the QCLE in conjunction with a transition filtering scheme (to improve convergence), where an average over $10^6$ trajectories fails to capture the exact long-time dynamics \cite{David.16.JCTC}.  In section II of the SI, we also consider the biased SBM (see section II of the SI for the full details of the model, equations of motion, and results).  As seen in figure 2 of the SI, in the low temperature regime, the DECIDE result exhibits quantitative deviations from the numerically exact one at long times, but captures the qualitative trend very well.  As explained in section II of the SI, this deviation is due to a pronounced memory effect at low temperatures in the biased SBM.  On the other hand, in the high temperature regime, DECIDE performs very well out to long times.

Now we turn to the PIET model \cite{Wang.04.CPL}, which has been previously used to study nonlinear spectroscopic signals related to PIET reactions in photosynthetic antenna complexes \cite{Vos.93.N,Michel.96.NULL,Engel.07.N,Panitchayangkoon.11.PNAS,Hwang.10.CM} and organic solar cells \cite{Park.09.NP}. The quantum subsystem is an ET complex with three electronic states:  a ground state $|g\rangle$, a photo-induced excited state $|d\rangle$ corresponding to the donor of the ET reaction, and an optically dark charge transfer state $|a\rangle$ corresponding to the acceptor of the ET reaction. The bath is composed of $N$ independent classical harmonic oscillators that are bilinearly coupled to the subsystem. The Hamiltonian of the total system is given by
\begin{eqnarray}
\hat{H}_W(t)&=& \sum_{m=g,d,a}\varepsilon_m|m\rangle\langle m|+\Delta(|d\rangle\langle a|+|a\rangle\langle d|)\nonumber\\
&&+\frac{1}{2}\sum_{j=1}^N\left[P_j^2+\omega_j^2\left(R_j+\frac{2C_j}{\omega_j^2}|a\rangle\langle a|\right)^2\right]\nonumber \\
&&-\mu E(t)(|g\rangle\langle d|+|d\rangle\langle g|),
\end{eqnarray}
where $\varepsilon_m$ is the site energy of $m$-th state, $\Delta$ is the donor-acceptor electronic coupling, $\mu$ is the transition dipole moment, and $E(t)=f_1(t-t_1)\cos[\omega_1(t-t_1)]$ is the incident laser field with frequency $\omega_1$ and Gaussian envelope $f_1(t-t_1)=\sqrt{\frac{4\ln 2}{\pi\tau_1^2}}\exp\left(-4\ln 2\frac{(t-t_1)^2}{\tau_1^2}\right)$ (which is centered at time $t_1$ and has a full-width at half-maximum (FWHM) $\tau_1$). The bilinear subsystem-bath coupling is characterized by a Debye-Drude spectral density $J(\omega)=\frac{\lambda_D}{2}\frac{\omega\omega_D}{\omega^2+\omega_D^2}$, where $\lambda_D$ is the bath reorganization energy and $\omega_D$ the characteristic frequency.

To monitor the progress of the PIET reaction following the photoexcitation by a laser pulse, we focus on the time-dependent population of the donor state, i.e., the expectation value of $\hat{\mathcal{P}}_{dd}=|d\rangle\langle d|$.  Therefore, an appropriate choice for the generalized coordinates of the subsystem is $\boldsymbol{\hat{x}}=(\{\hat{\mathcal{P}}_{mn}\})$, where $\hat{\mathcal{P}}_{mn}=|m\rangle\langle n|$ is the subsystem projection operator (with $m/n=g,d,a$). Given the condition that $\sum_{m=g,d,a}\hat{\mathcal{P}}_{mm}=1$, there are $9\times(8+2N)$ coupled FODEs for the matrix elements of the subsystem and bath coordinates in Eq.~(\ref{eq:eom_detail}). The initial density operator has the factorized form $\hat{\rho}_W(0)=\rho_{B,W}(0)\hat{\rho}_S(0)$, where $\hat{\rho}_S(0)=|g\rangle\langle g|$ and $\rho_{B,W}(0)$ has the same form as in the SBM.  We take $\{|\alpha\rangle\}=\{|g\rangle,|d\rangle, |a\rangle\}$; thus, according to Eq.~(\ref{eq:average}), the time-dependent population of the donor state is
\begin{equation}
\langle\hat{\mathcal{P}}_{dd}(t)\rangle~=~\int d\boldsymbol{X}(0)\rho_{B,W}(\boldsymbol{X}(0))\mathcal{P}_{dd}^{gg}(t),
\end{equation}
where we have used the fact that $\rho_S^{gg}(0)=1$.  

\begin{figure}[tbh!]
  \centering
  \includegraphics[width=0.9\columnwidth]{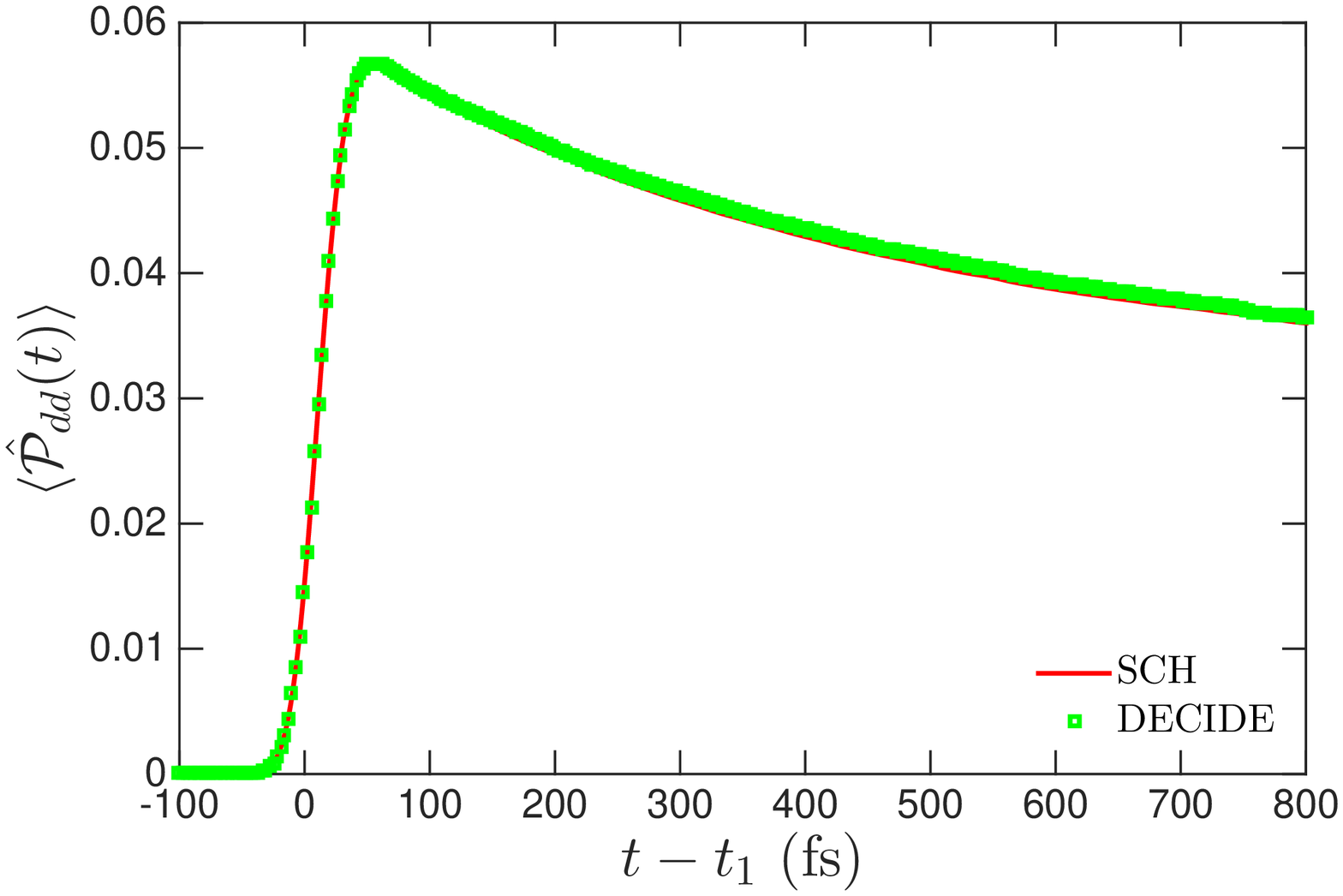}
\caption{Time evolution of $\langle\hat{\mathcal{P}}_{dd}(t)\rangle$ for the photo-induced electron transfer model. The parameters of the subsystem are $\varepsilon_g=0~\mathrm{cm}^{-1}$, $\varepsilon_d=13000~\mathrm{cm}^{-1}$, $\varepsilon_a=13000~\mathrm{cm}^{-1}$, and $\Delta=\mu=50\mathrm{cm}^{-1}$. The parameters of the bath are $T=300$ K, $\omega_D=50~\mathrm{cm}^{-1}$, $\lambda_D=500~\mathrm{cm}^{-1}$, $\omega_{max}=2500~\mathrm{cm}^{-1}$, and $N=40$. The laser pulse is centered at $t_1=100$ fs with a frequency $\omega_1=13000~\mathrm{cm}^{-1}$ and a FWHM $\tau_1=50$ fs. An ensemble of $1\times10^4$ trajectories and a MD time step $\Delta t=$1 fs were used to obtain our converged DECIDE result (green squares). The red solid line is the benchmark result calculated using the SCH method. }
\label{fig:pd}
\end{figure}

Our result for $\langle\hat{\mathcal{P}}_{dd}(t)\rangle$ is shown in figure \ref{fig:pd} (the simulation details can be found in section III of the SI). The benchmark result was obtained using the numerically exact self-consistent hybrid (SCH) method \cite{Wang.01.JCP,Wang.04.CPL}. As can be seen, the DECIDE results, generated using only $1\times10^4$ trajectories, are in excellent agreement with the benchmark result out to long times.  This result should be contrasted with that obtained by one of the authors using a surface-hopping solution of the QCLE, where an average over $3\times 10^7$ trajectories fails to exactly capture both the short- and long-time dynamics \cite{Rekik.13.JCP}.  

Finally, we consider the excitation energy transfer in the FMO complex, which can be described by a standard Frenkel exciton Hamiltonian in the single-excitation subspace \cite{Adolphs.06.BJ,Ishizaki.09.PNAS,Ke.16.JCP}
\begin{eqnarray}
\hat{H}_W &=& \sum_{n=1}E_n|n\rangle\langle n|+\sum_{m\neq n}V_{mn}|n\rangle\langle m|\nonumber\\
&&+\frac{1}{2}\sum_{n=1}\sum_{j=1}^M\left[P_{n,j}^2+\omega_{n,j}^2\left(R_{n,j}-\frac{C_{n,j}}{\omega_{n,j}^2}|n\rangle\langle n|\right)^2\right],
\end{eqnarray}
where $|n\rangle$ denotes the state of the $n$th chromophoric site with site energy $E_n$, and $V_{mn}$ is the excitonic coupling strength between the $n$th and $m$th site (the values of these parameters may be found in table 1 of the SI). Each site is coupled to an independent harmonic heat bath containing $M$ oscillators. The bilinear coupling to each bath is characterized by a Debye-Drude spectral density $J(\omega)=2\lambda_D\frac{\omega\tau_c}{1+\omega^2\tau_c^2}$, where $\lambda_D$ is the bath reorganization energy and $\tau_c$ the characteristic time.

As an illustration, we focus on the apo-FMO which contains seven bacteriocholorophyll (BChl) pigment-proteins per subunit (and the conventional numbering of the BChls has been used). Again, we choose the subsystem projection operators as the generalized coordinates for the subsystem, i.e., $\boldsymbol{\hat{x}}=(\{\hat{\mathcal{P}}_{mn}\})$. Given the condition that $\sum_{n}\hat{\mathcal{P}}_{nn}=1$, there are $49\times(48+2N)$ coupled FODEs for the matrix elements of the subsystem and bath coordinates in Eq.~(\ref{eq:eom_detail}), where $N=7M$. The initial density operator has the factorized form $\hat{\rho}_W(0)=\rho_{B,W}(0)\hat{\rho}_S(0)$, where $\hat{\rho}_S(0)=|1\rangle\langle 1|$ and $\rho_{B,W}(0)$ is the product of seven partially Wigner-transformed Gaussian distributions. We take $\{|\alpha\rangle\}=\{|1\rangle,|2\rangle,\cdots,|7\rangle\}$, thus, according to Eq.~(\ref{eq:average}), the time-dependent population for the $n$th chromophoric site is
\begin{equation}
\langle\hat{\mathcal{P}}_{nn}(t)\rangle~=~\int d\boldsymbol{X}(0)\rho_{B,W}(\boldsymbol{X}(0))\mathcal{P}_{nn}^{11}(t),
\end{equation}
where we have used the fact that $\rho_S^{11}(0)=1$.

\begin{figure}[tbh!]
  \centering
  \includegraphics[width=0.9\columnwidth]{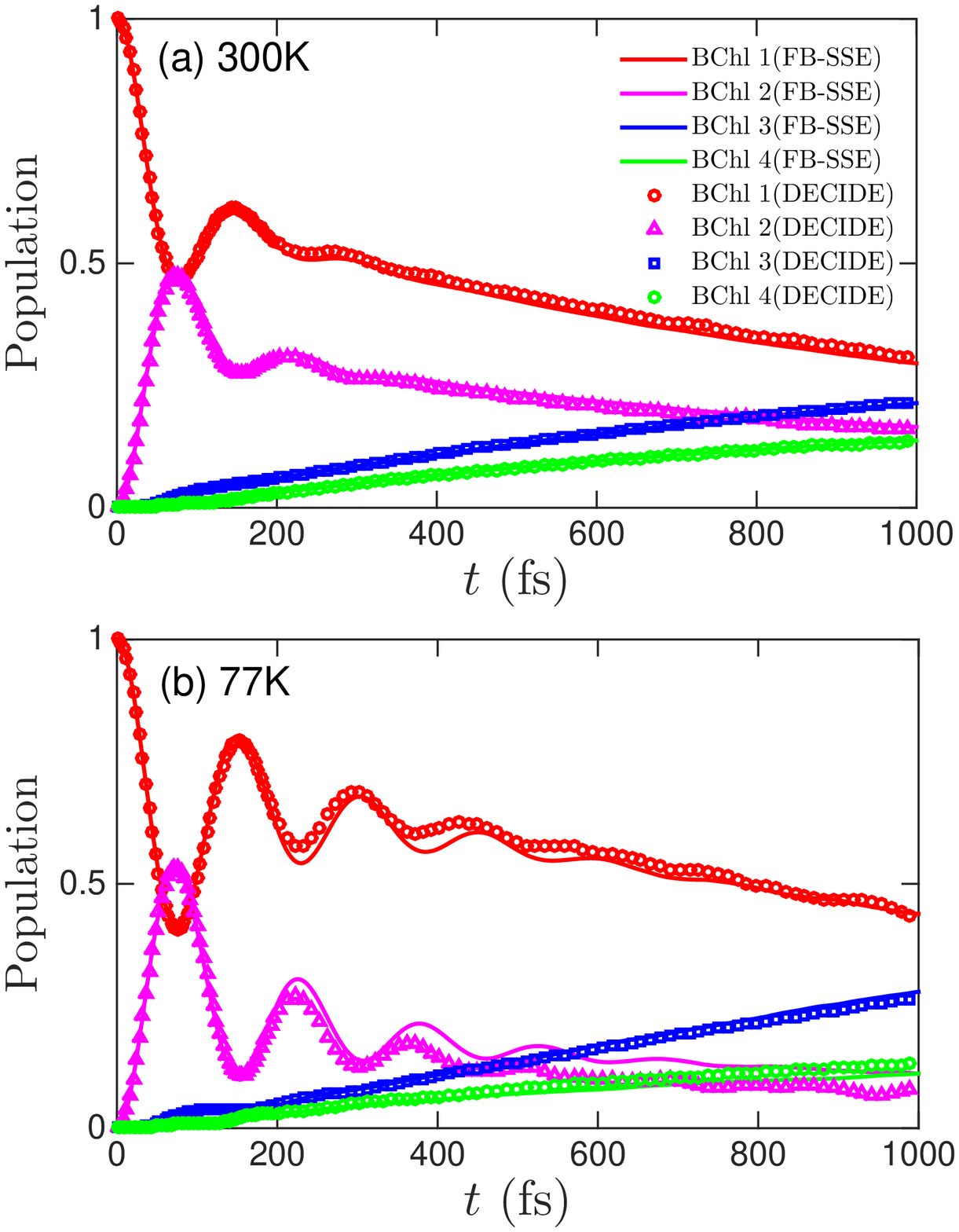}
\caption{Time evolution of the population of each BChl in the apo-FMO complex at (a) a physiological temperature (300 K) and (b) a cryogenic temperature (77 K). The parameters of the baths are $M=40$, $\tau_c=50$ fs, and $\lambda_D=35~\mathrm{cm}^{-1}$. An ensemble of $1\times10^4$ trajectories and a MD time step $\Delta t=$1 fs were used to obtain our converged DECIDE results (coloured shapes). The solid lines are the benchmark results calculated using the FB-SSH method.}
\label{fig:fmo}
\end{figure}

Our results for $\langle\hat{\mathcal{P}}_{nn}(t)\rangle$ at $T=$77 K and 300 K are shown in figure \ref{fig:fmo} (the simulation details can be found in section IV of the SI). The benchmark results were obtained using FB-SSE \cite{Ke.16.JCP}. We only present populations for the first four BChl pigments as the others are negligible.  
From the figure, we see that our method performs very well at both temperatures. It should be noted that these results were obtained with only $1\times 10^4$ trajectories, while the other QCLE-based methods used in previous studies of this model, namely the Poisson Bracket Mapping Equation (PBME) and Forward-Backward Trajectory Solution (FBTS), required at least two orders of magnitude more trajectories.  Although FBTS performed well at both temperatures\cite{Hsieh.13.JCP}, PBME gave rise to substantial deviations from the exact result at $77$ K \cite{Kelly.11.JPCL}.

In summary, we put forward a novel mixed quantum-classical dynamics method based on an approximate solution of the QCLE that does not involve surface-hopping.  Rather, this method involves solving a deterministic set of coupled FODEs for both the subsystem and bath coordinates expressed in an arbitrary basis (spanning the Hilbert space of the subsystem), and then constructing observables from the time-dependent coordinates.  Our results for the SBM, PIET, and FMO complex models considered in this study are in excellent agreement with those of the numerically exact approaches.  In contrast to the surface-hopping solutions of the QCLE, the current method requires several orders of magnitude fewer trajectories for convergence and is capable of generating highly stable long-time dynamics.  Owing to its favourable balance between accuracy and efficiency, the present method constitutes a powerful way of simulating the quantum dynamics of realistic systems.
 
 \begin{acknowledgements}
J. Liu would like to thank Yaling Ke for providing the FB-SSE results. We are also grateful to Dr.~Chang-Yu Hsieh, Prof.~Jeremy Schofield, and Prof.~Raymond Kapral for helpful comments and discussions. This work was supported by a grant from the Natural Sciences and Engineering Research Council of Canada (NSERC).
\end{acknowledgements}

\bibliography{a}

\end{document}